\documentstyle[11pt,aaspp4,amssym,flushrt]{article}

\lefthead{Cooray}
\righthead{Radio Emission from Lensed Arcs}

\begin{document}

\title{Radio Emission towards Lensed Arcs in Galaxy Clusters}

\author{Asantha R. Cooray\altaffilmark{1}}
\altaffiltext{1}{Department of Astronomy and Astrophysics, University of Chicago, Chicago IL 60637.}


\begin{abstract}

We present results from a search to detect
radio emission at 1.4 GHz from gravitationally
lensed arcs towards galaxy clusters. 
Using the FIRST survey
observations of galaxy cluster fields, 
we have clearly detected 1.4 GHz emission at the
location of bright arc towards
 the galaxy cluster Abell 370.
The possibility that this emission is associated with a 
projected galaxy superimposed on the arc cannot be completely
ruled out.
However, the detection of CO(2-1) line towards this arc
by Casoli {\it et al.} (1996) and the increase in 1.4 GHz
diffuse emission near the arc region
strongly suggest a lensed origin for detected radio emission.

\end{abstract}


\section{Introduction}

Since the discovery of giant arcs towards galaxy clusters Abell 370 and
CL 2244-02 (Lynds \& Petrosian 1996, Soucail {\it et al.} 1987), the search
for bright arcs towards galaxy clusters has steadily grown. 
These detections, now amounting to over 25 clusters with
at least one lensed arc, have allowed the construction of
galaxy cluster gravitational potentials responsible for
lensing and the study of high redshift galaxies which are magnified
through cluster 'gravitational telescopes'. We refer the
reader to Narayan \& Bartelmann (1996) 
for an overview of gravitational lensing in general, and Fort \& Mellier 
(1994) for a review of gravitational
lensing towards galaxy clusters. 

Currently, all of the known
arcs and arclets in galaxy clusters are optically selected.
This is contrary to the radio selected lenses which amount
to more than half the known gravitational lenses primarily 
due to galaxy-scale gravitational potentials. Although radio searches
have been carried out with the hope of detecting what may possibly
be called 'radio arcs' in galaxy clusters, no convincing evidence has yet
been found for such sources. Statistical studies of lensed radio emission
towards galaxy clusters have been controversial, with some studies
suggesting the detection of preferential tangential orientation,
as in due to a weak lensing effect (Bagchi \& Kapahi 1995), while
others suggesting no such evidence (e.g., Andernach {\it et al.} 1997).

The gravitational lensing of radio sources towards galaxy clusters 
can directly or indirectly affect important cosmological studies based
on galaxy clusters, such as observations of
the Sunyaev-Zeldovich (SZ) effect (Sunyaev \& Zeldovich 1970, 1980; see
Birkinshaw 1998 for a recent review). Loeb \& Refregier
(1997) suggested that lensed sources towards
clusters can produce an underestimation of the Hubble
constant as much as 13\% when derived 
by combining the SZ observations at 15 GHz
and X-ray emission observations. This is primarily due the fact that
when accounting for radio sources in SZ observations, 
there is more unaccounted radio flux towards the cluster
center region than outside regions, due to enhanced lensing
contribution towards center.
 In Cooray {\it et al.} (1998) we addressed the issue
of gravitationally lensed radio sources in clusters at 28.5 GHz based
on radio sources detected as part of the SZ observational program
using mm-arrays by Carlstrom {\it et al.} (1995).
In order to properly 
address the issue of radio contamination of SZ observations
and to
find direct evidence for lensed radio sources towards galaxy clusters,
we have initiated a study of X-ray luminous clusters at radio wavelengths.
As part of this study, we searched for radio emission 
from lensed gravitational
arcs in present radio data. Lensed emission from gravitational
arcs and images has been target of various studies
related to detection of molecular line emission at millimeter
wavelengths. Most of the CO lines in emission
have been detected in gravitationally lensed sources (e.g., Barvainis {\it
et al.} 1998), and Casoli {\it et al.} (1996) reported the detection of
 CO(2-1) line associated with the bright A0 arc towards
Abell 370 at a redshift of 0.725 (Soucail {\it et al.} 1988).
We restricted our search towards clusters where 
accurate astrometric information is available for optical
arcs. In order
to perform an unbiased radio search, we considered the two large area radio
surveys, mainly NVSS (Condon {\it et al.} 1998) and 
FIRST (Becker {\it et al.} 1995), using the VLA at 1.4 GHz. 
We opted to use the data from the VLA B-array 
FIRST survey, due to
the higher angular resolution and higher sensitivity of the
data 
when compared to the VLA D-array NVSS survey.

We present initial results from this search, which included
a radio search towards galaxy clusters Abell 370, Abel 963 and
Cl 2244-02. In Section 2 of the paper we present our results and
in Section 3 we discuss one strong detection of radio emission
associated with the bright arc in Abell 370. We also discuss
the possible sources of radio emission towards the arc
and show that the emission is more likely
due to lensing and is associated with the lensed
galaxy instead of a foreground source.

\section{Methods \& Results}

We searched the literature to find published astrometric locations of
strong arcs towards galaxy clusters which also overlap with the sky
coverage of FIRST data available as of early this year.
Even though there are over 20 clusters with lensed arcs, we failed
to find astrometric information for over 40\% of them in published papers
and for over 15\% of the clusters we found that
the accuracy of published
coordinates was less than what was required for the present
radio counterpart search.
Also, the limited area of the sky for which calibrated FIRST survey
 data was available restricted our search to 3 clusters: Abell 370,
Abell 963, and Cl 2244-02. We refer the reader to Becker {\it et al.} (1995)
 for details
on FIRST survey, including observational and data reduction techniques.

In Table 1, we provide observational parameters
of the arcs studied as part of this search, including 
astrometric and redshift information for them and the clusters towards
which they are detected. We searched the regions near these arcs
in the final calibrated FIRST survey image data. We expect that the uniform
reduction and calibration of raw data will not bias our search from one 
cluster field to another. In Table 2, we  list the observed 1.4 GHz
flux density, both peak and integrated flux densities, 
and rms noise. As tabulated, we have clearly detected the 1.4 GHz emission
towards the lensed arc A0 in Abell 370. The peaked emission
is $\sim$ 1.9$''$ away from the tabulated optical position
of the arc from Smail {\it et al.} (1993). This offset is roughly equivalent
to 1 pixel (1.8$''$) in the FIRST image used for the analysis, and
could be improved with better optical astrometric information. 
However, given that the optical arc is extended over 20$''$, 
this small offset does not by itself rule out
that the emission is associated with the lensed arc.
Also, up to 1.5$''$ offset is expected between the radio and optical
astrometric frames, which needs to be improved by combining
preferably astrometric information of stellar objects
 which are bright in radio.

In Fig. 1, we show the region associated with the gravitational
lensed arcs towards Abell 370. The circularly marked regions
are the expected central locations of
two arcs, and as shown, bright emission is detected towards
the northern arc A0 with an integrated flux density of 1.48 mJy and rms
noise level of
 0.16 mJy beam$^{-1}$. The peaked emission coincides with
the lensed optical arc within 1.9 arcseconds. The bright
emission can be fitted with an elliptical Gaussian model of 
7.4$''$ by 6.8$''$ and a positional
angle of 91 degrees.  According to the FIRST catalog, 
the observed emission is
 best fitted with an ellipse of 4.9$''$ by 1.7$''$
and a positional angle of 90 degrees based on the
deconvoluted elliptical point-spread function. 
The surrounding region
of this peaked 
source shows a substantial increase in diffuse emission, especially
to the east. However, this diffuse
emission
is not clearly imaged with the existing B-array data. The region
surrounding lensed arc
A5 does not show any evidence for 1.4 GHz radio emission
at the present sensitivity level,  with an upper limit
on 1.4 GHz integrated flux density of 0.55 mJy and rms 0.17 my beam$^{-1}$.

In Fig. 2, we show the 1.5$'$ by 1.5$'$ region centered on the
bright arc towards galaxy cluster Cl2244-02. Here again, we do not
find evidence of radio emission. The bright radio source to the west 
is TXS2233-022 (Douglas {\it et al.} 1996) 
with a peak flux density of  10.6 mJy at 1.4 GHz. The upper limit
on the lensed arc region is 0.66 mJy with a rms noise of 0.18 mJy beam$^{-1}$.
Also, we did not find any evidence for radio emission associated with
the bright northern arc towards Abell 963, with an upper limit on
the integrated flux density of 0.45 mJy 
with a rms noise of 0.20 mJy beam$^{-1}$.

\section{Discussion}

The observed data clearly suggest 1.4 GHz radio emission towards the
the gravitationally lensed  arc A0 in Abell 370. We looked
for literature and databases to find radio emission around this region
at other centimeter wavelengths, but failed to find any.
The upper limit on the 28.5 GHz radio emission around
this region is 0.32 mJy with a rms noise of 0.12 mJy beam$^{-1}$ (Cooray
{\it et al.} 1998), but the large synthesized beam of
these observations, targeted at imaging the SZ effect,
do not allow the flux density to be estimated accurately.
Also, out of the three clusters studied in this paper,
Abell 370 is the only cluster 
in the Cooray {\it et al.} (1998) sample of clusters selected for
SZ observations at OVRO and BIMA arrays outfitted with cm-wave
receivers.

The observed 1.4 GHz 
emission is clearly peaked, instead of an arc-like emission
as one would expect, if the radio emission is lensed
and follows optical emission. However, if the radio emission is 
due to a smaller region of the lensed galaxy, such as a nuclear region,
 than the optical emission,
then this peak can easily be explained even with gravitational
lensing. This situation is also similar to some examples
of known gravitational lenses where optical and infrared
observations show the existence of a ring, while radio
data only shows partial peaked emission at different locations of
the ring (e.g., King {\it et al.} 1997).

The arc A0 in Abell 370
extends 20$''$ with an estimated width of $\sim$ 1.5$''$ (see Fig. 1
in Kneib {\it et al.} 1994). The spectral energy distribution (SED) 
of this arc between 2000 and 16000 $\AA$ 
is well fitted  by a normal spiral galaxy of class Sbc (Aragon-Salamanca
\& Ellis 1990), where as the SEDs of higher redshift A5 arc and
the arc in Cl2244-02 suggest the
existence of young stars, closely matching the SED of an intense
star forming system (Smail {\it et al.} 1993).
The non detected arcs are either at high redshift (2.237 for CL 2244-02)
or have a less surface brightness when compared with the bright 
arc in Abell 370.
The magnification factor for arc A0 as derived by detailed
lens modeling by Kneib {\it et al.} (1992) is 14. As estimated
by Casoli {\it et al.} (1996), the arc corresponds to an unlensed galaxy of
size 3$'' \times 1''$. As shown in Fig. 1, the radio emission is
clearly unresolved in the existing B-array data. Further observations
with A-array should allow better mapping of the emission, and
is requird to test the lensing hypothesis presented here.

It is likely that the radio emission is associated with the
nuclear region of the lensed galaxy. However, an estimate of the
spectral index for radio emission can aid in the identification of
emission region, or at least constraining certain radio emission
mechanisms. The limit of radio emission at 28.5 GHz do not
allow an accurate measurement of the spectral index, and observations
perhaps at 4.85 GHz using the VLA A-array are much needed.
If a steep-spectrum is found, then this emission
may be associated with an AGN-related emission. However, given
that the arc is a lensed spiral galaxy, strong AGN activity is
not expected. The radio emission can also be due to star formation,
either ongoing or recent cessation of star formation, but
Smail {\it et al.} (1993) finds that this arc at a redshift of 0.725
doesn't show enough evidence to suggest intense star formation.
However, the radio emission can be directly due to supernova
remnants, as suggested by Helou {\it et al.} (1985), and the
flux density is in agreement with their estimates of supernova
rates in spiral galaxies. 

Apart from lensing hypothesis for radio emission, 
there is also an alternative and obvious explanation to
radio emission, which is foreground emission imposed on the
region of lensed optical arc. We can estimate the probability
for positional coincidence based on the observed number of
optical galaxies and
radio sources in the field of Abell 370. In a FIRST image of
10$'$ by 10$'$ centered on the Abell 370 cluster center, we find
14 radio sources with peak flux density greater than 1 mJy.
In the 1$'$ by 1$'$ region centered on the lensed arc, we
find close to 25 optical sources with V magnitude greater than
22. Assuming that the radio emission center extends over a region of
5$''$ by 5$''$ and allowing a 1.5$''$ error for any difference
between optical and radio astrometric frames, the probability
that the radio emission is due to a source other than arc is
close to 10$^{-3}$. It is more likely that radio emission is
associated with the lensed arc rather than a foreground galaxy.
However, if optical sources down to 25th magnitude is considered
this probability increases substantially, but with the observed
increase in diffuse brightness surrounding the optical arc,
it is not possible to rule out gravitational lensing immediately.
A reason for diffuse emission is required and it may well be
associated with gravitational lensing. 
An alternative explanation is that this emission is
realted to a cluster wide diffused halo type
source, such as the ones 
observed towards Coma (e.g., Feretti \& Giovannini 
1998). However, such
halo sources are expected to brighten near the central region of
the cluster instead of 30$''$ to 40$''$ away form center where arcs
are observed at the distance of typical Einstein radii.
Therefore, reasons for diffuse emission as associated with cluster
electrons can only be understood only if a possible 
galaxy cluster merger and
subcluster at the location of the bright arc
is introduced. 
Simple interpretation of the diffuse emission as due to
gravitational lensing needs to be further investigated and
sensitive observations perhaps will enable the modeling
of such emission based on gravitational lensing. Such
an analysis has been performed to explain the low brightness
surface emission observed in the field containing gravitational
lens 0957+561 by Avruch {\it et al.} (1997).

It is expected that
more radio observational work will be carried out
near this arc region to map both the radio emission towards
the arc and the diffuse emission surrounding it
at a resolution better than what is currently available.
Also, in the optical side
more astrometric work is required to
obtain enough positional accuracy to do a better radio counterpart
search of lensed arcs than that it performed here with current
information. Until then, we cannot rule out the possibility that
the observed radio emission is associated with the lensed galaxy.
If lensing interpretation is true, then the arc A0 in abell 370
 is the first arc to show radio emission of a lensed arc towards
a galaxy cluster.

\acknowledgments

ARC gratefully acknowledges useful discussions with John Carlstrom, Andre
Fletcher, Laura Grego, and
Bill Holzapfel.

\clearpage

\figcaption[fig1.ps]{Gray scale image of the VLA B-array NVSS survey
observations of Abell 370. Shown here is a 1.5$'$ by 1.5$'$ field
centered on the gravitationally lensed arc A0. The radio emission
is marked by a circle, and the expected location of the arc A5 is marked
to the south of it with another circle. The noise level in this
map is about 160 $\mu$Jy beam$^{-1}$, and the pixel size is 1.8$''$.
\label{fig1}}

\figcaption[fig2.ps]{Gray scale image of the VLA B-array NVSS survey
observations of Cl2244-02. Shown here is a 1.5$'$ by 1.5$'$ field
centered on the gravitationally lensed arc. The expected location
of the radio emission
is marked by a circle, and the bright source to the
left of this circle has a peaked 1.4 GHz flux density of 10.6 mJy.
\label{fig2}}
 
\clearpage

\begin{deluxetable}{lllll}
\tablewidth{30pc}
\tablenum{1}
\tablecaption{Optical observational parameters}
\tablehead{
\colhead{Arc}    &
\colhead{$\alpha$ (J2000.0) \tablenotemark{1}} & 
\colhead{$\delta$ (J2000.0) \tablenotemark{1}} &
\colhead{z$_{\rm cl}$} & \colhead{z$_{\rm arc}$} }
 
\startdata
A370-A0 & 02 39 52.9 & -01 34 59 & 0.375 & 0.725\tablenotemark{2} \nl
A370-A5 & 02 39 53.3 & -01 35 24 & 0.375 & 1.308\tablenotemark{3} \nl
A963-N  & 10 16 55.7 &  39 02 18 & 0.206 & 0.771\tablenotemark{4} \nl
Cl2244-02 & 22 47 12.9 & -02 05 40 & 0.329 & 2.238\tablenotemark{3} \nl
\enddata
\tablenotetext{1}{Smail {\it et al.} 1993.}
\tablenotetext{2}{Soucail {\it et al.} 1988.}
\tablenotetext{3}{Mellier {\it et al.} 1991.}
\tablenotetext{4}{Ellis {\it et al.} 1991.}
\end{deluxetable}

\begin{deluxetable}{lllccc}
\tablewidth{40pc}
\tablenum{2}
\tablecaption{1.4 GHz Radio Detections}
\tablehead{
\colhead{Arc}    &
\colhead{$\alpha$ (J2000.0) \tablenotemark{1}} & 
\colhead{$\delta$ (J2000.0) \tablenotemark{1}} &
\colhead{S$_{\rm peak}$ (mJy)} & 
\colhead{S$_{\rm int}$ (mJy)} & 
\colhead{S$_{\rm rms}$ (mJy beam$^{-1}$)}}  
\startdata
A370-A0 & 02 39 52.97 & -01 35 00.6 & 1.02 & 1.48 & 0.16 \nl
A370-A5 & 02 39 53.35 & -01 35 24.5 & $<$0.45  & $<$0.55 & 0.17 \nl
A963-N  & 10 16 55.75 &  39 02 18.5 & $<$0.38 & $<$0.45 & 0.20 \nl
Cl2244-02 & 22 47 12.95 & -02 05 40.5 & $<$ 0.52 & $<$0.66 & 0.18 \nl
\enddata
\tablenotetext{1}{This is the location of peaked emission.}
\end{deluxetable}

\begin{figure}
\plotone{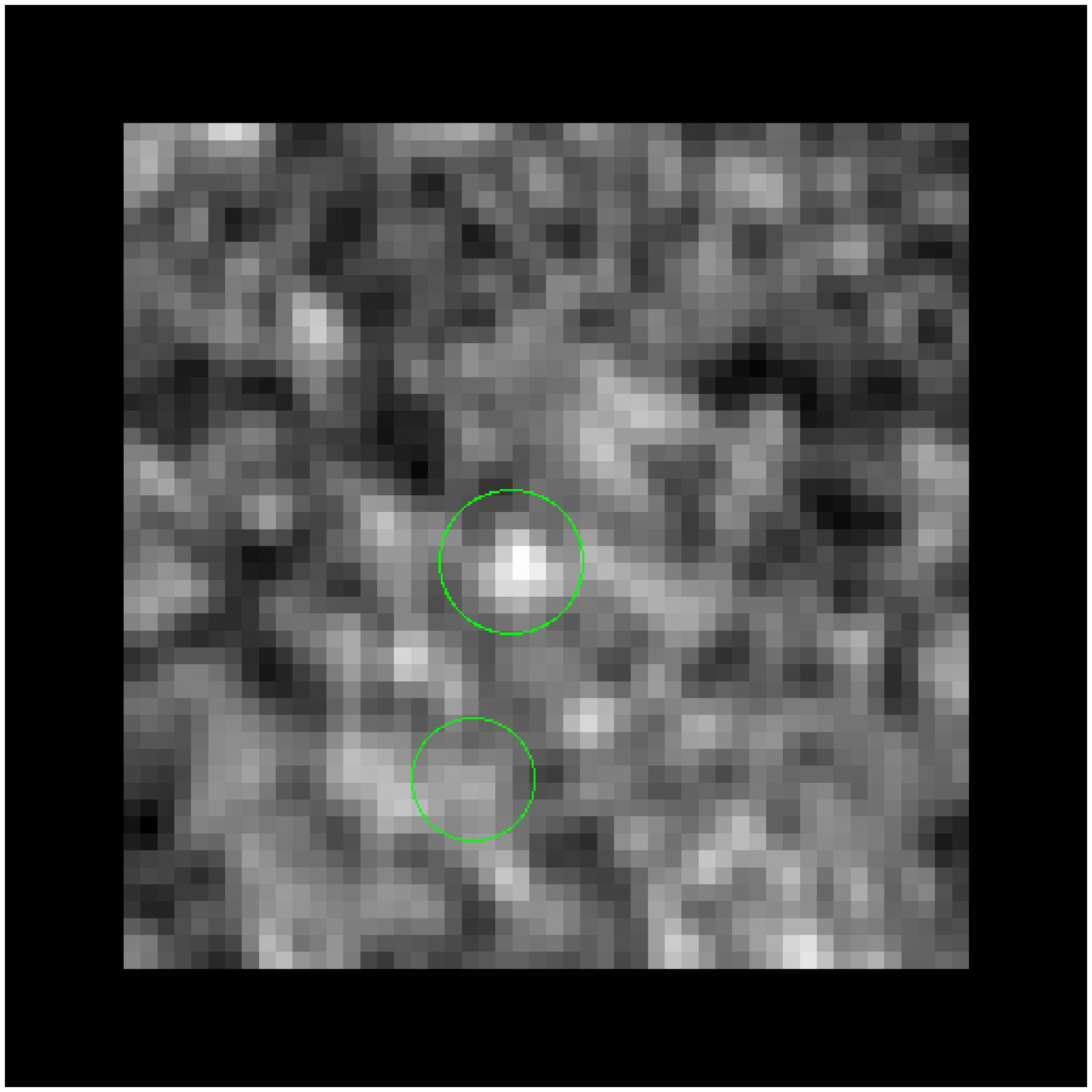}
\end{figure}

\begin{figure}
\plotone{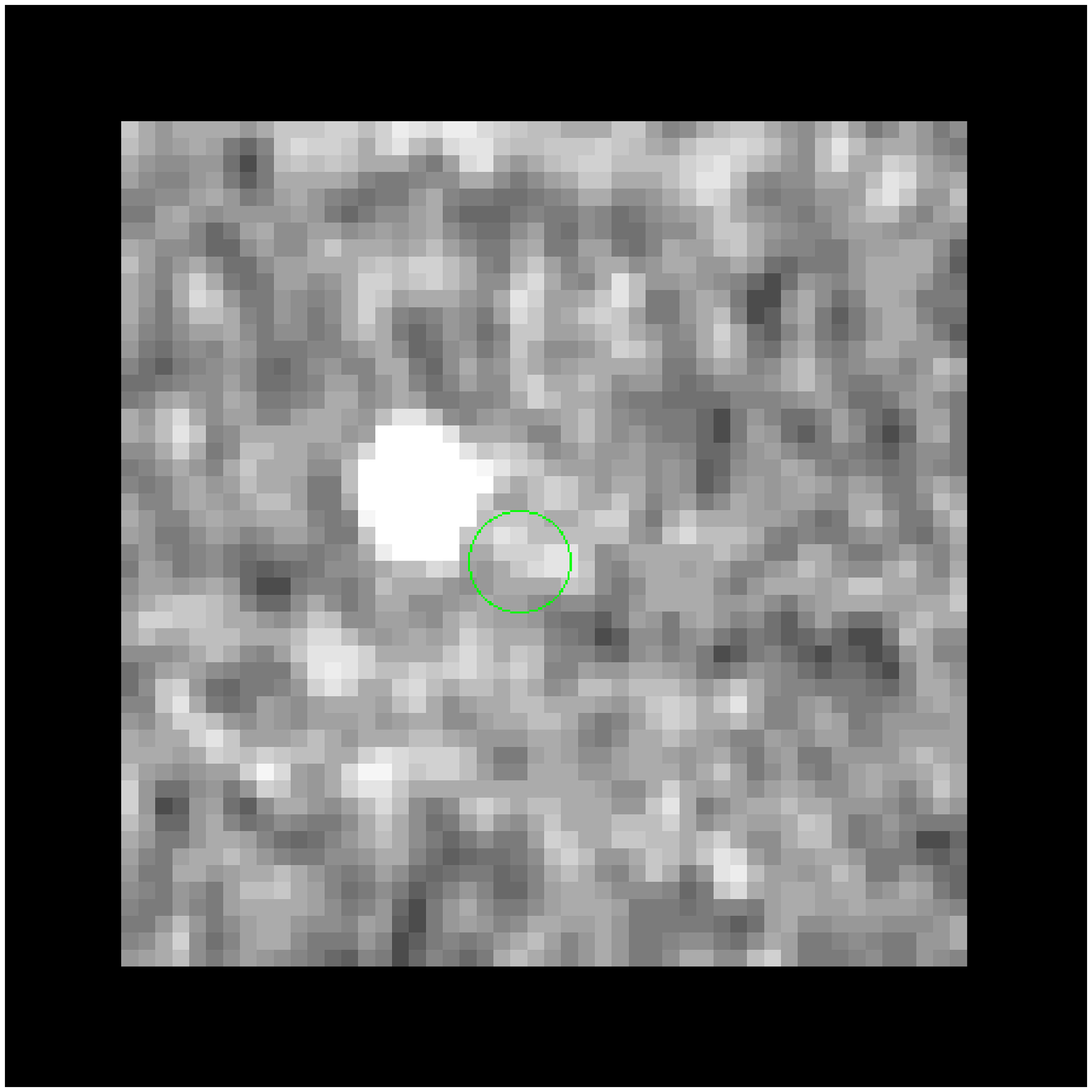}
\end{figure}

\end{document}